\documentclass[11pt,a4paper]{article}

\usepackage[utf8]{inputenc}
\usepackage[T1]{fontenc}
\usepackage{lmodern}
\usepackage{amssymb,amsthm,amsfonts,mathrsfs}
\usepackage{geometry}
\usepackage{mathtools}
\usepackage[all]{xy}
\usepackage{graphicx}
\usepackage{microtype}
\usepackage{enumitem}
\usepackage{stmaryrd}
\usepackage{amscd}
\usepackage[hidelinks]{hyperref}

\theoremstyle{plain}
\newtheorem{theorem}{Theorem}[section]
\newtheorem{lemma}[theorem]{Lemma}
\newtheorem{proposition}[theorem]{Proposition}

\newtheorem*{proposition*}{Proposition}
\newtheorem*{theorem*}{Theorem}

\theoremstyle{definition}
\newtheorem{definition}[theorem]{Definition}
\newtheorem{example}[theorem]{Example}
\newtheorem{notation}[theorem]{Notation}

\theoremstyle{remark}
\newtheorem{remark}[theorem]{Remark}

\newcommand{\Z}{\mathbb{Z}}
\newcommand{\R}{\mathbb{R}}
\newcommand{\C}{\mathbb{C}}
\newcommand{\N}{\mathbb{N}}
\newcommand{\Q}{\mathbb{Q}}
\newcommand{\K}{\mathbb{K}}

\DeclareMathOperator{\dist}{dist}
\DeclareMathOperator{\id}{id}
\DeclareMathOperator{\Hom}{Hom}

\DeclareMathOperator{\mesh}{mesh}
\DeclareMathOperator{\vol}{vol}
\DeclareMathOperator{\Vol}{Vol}

\DeclareMathOperator{\im}{im}
\DeclareMathOperator{\Tor}{Tor}
\DeclareMathOperator{\Diff}{Diff}
\DeclareMathOperator{\free}{free}
\DeclareMathOperator{\sd}{sd}
\DeclareMathOperator{\curv}{curv}
\DeclareMathOperator{\Curv}{Curv}

\title{Simplicial Cheeger--Simons models and simplicial higher abelian gauge theory}

\author{Jyh-Haur Teh\thanks{Department of Mathematics, National Tsing Hua University, Taiwan.
E-mail: \texttt{jyhhaur@math.nthu.edu.tw}.}}
\date{}

\begin{document}
\maketitle

\begin{abstract}
A pair $(K,K')$ consisting of a smooth triangulation $K$ of a compact smooth oriented Riemannian manifold $M$ and a sufficiently fine subdivision $K'$ determines a finite-dimensional Cheeger--Simons model $\mathscr{CS}(K,K')$ built from Whitney-type data on the induced curvilinear complexes. Its associated differential character groups $\Diff^{\bullet}(\mathscr{CS}(K,K'))$ provide a simplicial, finite-dimensional counterpart of the Cheeger--Simons differential characters $\widehat H^{\bullet}(M)$. We prove that every smooth triangulation admits a subdivision $K'$ for which $(K,K')$ is a Cheeger--Simons triangulation in this sense. Under a uniform fullness (shape-regularity) hypothesis, we show that the natural discretization/extension maps between $\widehat H^{k}(M)$ and $\Diff^{k}(\mathscr{CS}(K,K'))$ approximate the identity in a Sobolev-dual seminorm as $\mesh(K')\to 0$. For closed $M$, we further identify $\widehat H^{k}(M)$ canonically with the inverse limit of $\Diff^{k}(\mathscr{CS}(K,K'))$ over refinements. As an application, we formulate a simplicial higher abelian gauge theory whose gauge-invariant configuration space is $\Diff^{p}(\mathscr{CS}(K,K'))$, and we prove that the resulting simplicial (regularized) partition function converges, in the refining limit, to the corresponding smooth regularized partition function of Kelnhofer.
\end{abstract}

\section{Introduction}

Differential cohomology provides a bridge between topology and geometry: it refines an integral cohomology class by remembering differential-form data (curvature) together with secondary information such as holonomy.
Among the classical concrete models are the Cheeger--Simons differential characters \cite{CS, HLZ1, HL1, HL2, SS}, whose groups $\widehat H^k(M)$ fit into fundamental exact sequences intertwining
$H^k(M;\Z)$, de~Rham forms with integral periods, and tori of flat classes.
From the perspective of quantum field theory, these groups are natural configuration spaces for generalized abelian and higher-form gauge fields, where topological sectors (quantized fluxes) and geometric data (curvature representatives) are simultaneously essential; see for example \cite{KelnhoferGribov,K2,S}.

A persistent obstacle for rigorous constructions is analytic: smooth differential character groups are infinite-dimensional in the directions coming from differential forms, while functional integrals in physics are formal and require regularization.
One strategy is to replace smooth data by combinatorial or simplicial models that retain the relevant topological content but have finite-dimensional ``form parts''.
In higher abelian gauge theory this philosophy is especially effective because the gauge-inequivalent configuration space admits a description in terms of harmonic data plus coexact directions, leading to explicit determinant--torsion--theta-function structures in the regularized partition function \cite{K2}.

This paper develops an axiomatic and functorial framework for differential characters that applies both in smooth and in discrete settings, and uses it to build a simplicial approximation of differential characters via Whitney forms on refinements of a smooth triangulation.
We then apply these discrete models to define a finite-dimensional simplicial higher abelian gauge theory whose partition function converges to the smooth one.

\medskip
\noindent\textbf{1. Cheeger--Simons models and spark groups.}
We introduce \emph{Cheeger--Simons models}: triples $(E^\bullet,I_\bullet,\int)$ consisting of a cochain complex, a free chain complex, and a nondegenerate pairing satisfying Stokes and a de~Rham-type quasi-isomorphism.
To each model we associate differential character groups $\Diff^k(E^\bullet,I_\bullet,\!\int)$, defined in exact analogy with the classical Cheeger--Simons definition, and we prove that they satisfy the standard short exact sequences relating curvature, characteristic class, and flat data.
We further relate Cheeger--Simons models to Harvey--Lawson spark complexes by constructing an associated spark complex and proving a natural isomorphism
\[
\Diff^k(\mathscr{CS}) \cong \widehat H^k(\mathscr{S}),
\]
allowing structural results to be transferred between the two frameworks.

\medskip
\noindent\textbf{2. Whitney models from triangulations and approximation of $\widehat H^k(M)$.}
Let $M$ be a compact, oriented manifold without boundary and let $\chi:|K|\to M$ be a smooth triangulation.
After passing to a suitable subdivision $K'$, we construct a Cheeger--Simons model $\mathscr{CS}(K,K')$ from Whitney forms on an associated curvilinear subdivision.
Along sequences of refinements with \emph{uniform fullness} (i.e.\ nondegenerate simplex shapes) and $\mesh(K')\to 0$, we show that the natural comparison maps between smooth and simplicial characters become asymptotically inverse in a Sobolev-type topology; see Theorem~\ref{thm:Diff_converges}.

\medskip
\noindent\textbf{3. Simplicial higher abelian gauge theory and convergence of partition functions.}
For each Cheeger--Simons triangulation $(K,K')$ and degree $p\ge 0$, we take $\Diff^{p}(\mathscr{CS}(K,K'))$ as the gauge-invariant configuration space of a simplicial higher abelian gauge theory.
We formulate the action and observables so that they are compatible with the smooth theory under the comparison maps.
Under uniform fullness and consistency of discretization across topological sectors, we prove that the resulting simplicial partition functions converge to the smooth (regularized) partition function, in parallel with Kelnhofer's general formula and its determinant/torsion structure \cite{K2}; see Theorem~\ref{thm:Z_converges}.

\medskip
\noindent\textbf{4. Inverse limits.}
We prove an inverse-limit identification
\[
\varprojlim_{(K,K')}\Diff^k(K,K') \cong \widehat H^k(M)
\]
over the directed system of Cheeger--Simons triangulations, making precise in what sense the discrete models recover the smooth differential character group; see Theorem~\ref{thm:invlim-diff}.

\medskip
\noindent\textbf{Organization.}
Section~\ref{sec:CS_models} develops the axiomatic notion of Cheeger--Simons models and the comparison with spark complexes.
Section~\ref{sec:simplicial_CS} constructs Whitney-type Cheeger--Simons models from triangulations and proves approximation results.
Section~\ref{sec:gauge} formulates the simplicial higher abelian gauge theory and establishes convergence of partition functions.
Section~\ref{sec:invlim} proves the inverse-limit identification of triangulated differential characters with $\widehat H^k(M)$.
Section~\ref{sec:examples} contains worked examples and computations.

\section{Cheeger--Simons models and spark groups}\label{sec:CS_models}

\subsection{Cheeger--Simons models}

\begin{definition}[Cheeger--Simons model]
Let $I_{\bullet}=(\{I_k\}_{k\ge 0},\partial)$ be a chain complex and
$E^{\bullet}=(\{E^k\}_{k\ge 0},d)$ a cochain complex.
Assume that for each $k\ge 0$ we are given a pairing
\[
\int: I_k\times E^k \longrightarrow \K,\qquad (\alpha,\omega)\longmapsto \int_{\alpha}\omega .
\]
We say that $(E^{\bullet},I_{\bullet},\int)$ is a \emph{Cheeger--Simons model} over $\K$ if:
\begin{enumerate}[leftmargin=2em]
\item $I_{\bullet}$ is a chain complex of free abelian groups.
\item The pairing $\int$ is bilinear and nondegenerate.
\item (\emph{No nonzero integral-period elements}) For $k>0$, if $\omega\in E^{k}$ satisfies
$\int_{\beta}\omega \in \Z$ for all $\beta\in I_{k}$, then $\omega=0$.
\item (\emph{Stokes}) For $\alpha\in I_{k+1}$ and $\omega\in E^k$,
\[
\int_{\partial \alpha}\omega=\int_{\alpha}d\omega .
\]
\item (\emph{de Rham}) The map $H^k(E^{\bullet}) \to H^k(\Hom(I_{\bullet},\K))$ induced by
$\omega\mapsto \left(\alpha\mapsto \int_\alpha \omega\right)$ is an isomorphism.
\end{enumerate}
\end{definition}

\begin{example}
Let $M$ be a smooth compact oriented manifold without boundary,
$E^k=\Omega^k(M)$ the smooth differential $k$-forms, and $I_k=I_k(M)$ the group of smooth singular
$k$-chains with integral coefficients. With the usual integration pairing, $(\Omega^{\bullet}(M),I_{\bullet}(M),\int)$
is a Cheeger--Simons model over $\K=\R$.
\end{example}

\begin{notation}
Let $(C_{\bullet},\partial)$ be a chain complex and $(E^{\bullet},d)$ a cochain complex.
We write
\[
Z_k(C_{\bullet})=\ker(\partial:C_k\to C_{k-1}),
\qquad
Z^k(E^{\bullet})=\ker(d:E^k\to E^{k+1}).
\]
For a Cheeger--Simons model $(E^{\bullet},I_{\bullet},\int)$ we set
\[
I^k:=\Hom(I_k,\Z),\qquad F^k:=\Hom(I_k,\K),
\]
and we denote by $\Phi:I^{\bullet}\hookrightarrow F^{\bullet}$ the inclusion and by
$\Phi_*:H^k(I^{\bullet})\to H^k(F^{\bullet})$ the induced homomorphism.
The cochain differential on $F^{\bullet}$ (and on $I^{\bullet}$) will be denoted by $\delta$:
\[
(\delta T)(\beta):=T(\partial \beta),\qquad T\in F^k,\ \beta\in I_{k+1}.
\]
\end{notation}

\begin{definition}
The \emph{$k$-th differential character group} of a Cheeger--Simons model
$\mathscr{CS}=(E^{\bullet},I_{\bullet},\int)$ is
\begin{align*}
\Diff^k(\mathscr{CS})
=\Bigl\{ f\in \Hom(Z_k(I_{\bullet}),\K/\Z)\ \Big|\ &\exists\, \omega\in E^{k+1}\ \text{such that}\\
& f(\partial \beta)\equiv \int_{\beta}\omega \ \ (\mathrm{mod}\ \Z),\ \forall\,\beta\in I_{k+1}\Bigr\}.
\end{align*}
The element $\omega$ is called a \emph{curvature representative} (or \emph{form part}) of $f$.
\end{definition}

\medskip
For the rest of this subsection, fix a Cheeger--Simons model $\mathscr{CS}=(E^{\bullet},I_{\bullet},\int)$.

\begin{lemma}\label{lem:vanish_on_cycles}
Let $G$ be a divisible abelian group. If $g\in \Hom(I_k,G)$ satisfies $\delta g=0$ and $g|_{Z_k(I_{\bullet})}=0$, then
$g=\delta g'$ for some $g'\in \Hom(I_{k-1},G)$.
\end{lemma}

\begin{proof}
Since each $I_k$ is free abelian and $G$ is divisible, the universal coefficient theorem gives
$H^k(\Hom(I_{\bullet},G))\cong \Hom(H_k(I_{\bullet}),G)$ (see \cite[Thm.~7.61]{Rot}).
The class $[g]$ corresponds to the homomorphism $H_k(I_{\bullet})\to G$ induced by $g|_{Z_k}$.
If $g|_{Z_k}=0$, then $[g]=0$, hence $g$ is a coboundary: $g=\delta g'$ for some $g'\in \Hom(I_{k-1},G)$.
\end{proof}

\begin{proposition}\label{prop:delta_maps_exist}
Assume that the coefficient group $\K$ is divisible (as an abelian group), for example $\K=\R$.
Then there are surjective homomorphisms
\[
\delta_1:\Diff^k(\mathscr{CS})\to E^{k+1}_0,\qquad
\delta_2:\Diff^k(\mathscr{CS})\to H^{k+1}(I^{\bullet}),
\]
where
\[
E^{k+1}_0:=\Bigl\{\omega\in E^{k+1}\ \Big|\ d\omega=0,\ \int_{\alpha}\omega\in\Z\ \forall\,\alpha\in Z_{k+1}(I_{\bullet})\Bigr\}.
\]
\end{proposition}

\begin{proof}
Let $f\in\Diff^k(\mathscr{CS})$. Choose an extension $\widetilde{f}:I_k\to \K/\Z$ of $f$.
Then $\K/\Z$ is divisible and hence injective as a $\Z$-module, so such extensions exist.
Because $I_k$ is free, the quotient map $q:\K\to \K/\Z$ admits a lift $T:I_k\to \K$ with $q\circ T=\widetilde{f}$.

By definition of $\Diff^k$, there exists $\omega\in E^{k+1}$ such that
\[
q\bigl((\delta T)(\beta)\bigr)=\widetilde{f}(\partial \beta)=f(\partial\beta)
\equiv \int_{\beta}\omega \ \ (\mathrm{mod}\ \Z)\qquad \forall\,\beta\in I_{k+1}.
\]
Let $\widetilde{\omega}\in F^{k+1}=\Hom(I_{k+1},\K)$ denote the image of $\omega$ under the injective map
$E^{k+1}\hookrightarrow F^{k+1}$ induced by the pairing.
Then there exists $c\in I^{k+1}=\Hom(I_{k+1},\Z)$ such that
\[
\delta T=\widetilde{\omega}-c.
\]
Applying $\delta$ gives $\delta\widetilde{\omega}=\delta c$.
By Stokes, $\delta\widetilde{\omega}=\widetilde{d\omega}$, hence for every $\gamma\in I_{k+2}$ we have
\[
\int_{\gamma} d\omega \;=\; (\delta\widetilde{\omega})(\gamma)\;=\;(\delta c)(\gamma)\in \Z.
\]
By axiom (3) in degree $k+2$, it follows that $d\omega=0$, hence $\delta c=0$.
Moreover, for $\alpha\in Z_{k+1}(I_{\bullet})$ we have
\[
\int_{\alpha}\omega=\widetilde{\omega}(\alpha)=\delta T(\alpha)+c(\alpha)=c(\alpha)\in\Z,
\]
so $\omega\in E^{k+1}_0$.

Define $\delta_1(f):=\omega$ and $\delta_2(f):=[c]\in H^{k+1}(I^{\bullet})$.
Independence of choices follows from Lemma~\ref{lem:vanish_on_cycles} by the usual lift-change argument.

Surjectivity of $\delta_1$: given $\omega\in E^{k+1}_0$, the cochain $q\circ\widetilde{\omega}\in \Hom(I_{k+1},\K/\Z)$
vanishes on cycles and hence represents the zero class in $H^{k+1}(\Hom(I_{\bullet},\K/\Z))$. Thus
$q\circ\widetilde{\omega}=\delta t$ for some $t\in \Hom(I_k,\K/\Z)$. Lift $t$ to $\widetilde{t}:I_k\to\K$ and set $T:=\widetilde{t}$.
Then $\delta T=\widetilde{\omega}-c$ for some $c\in I^{k+1}$, and $f:=t|_{Z_k}$ satisfies $\delta_1(f)=\omega$.

Surjectivity of $\delta_2$: given $[c]\in H^{k+1}(I^{\bullet})$, choose $\eta\in F^{k+1}$ representing $\Phi_*[c]\in H^{k+1}(F^{\bullet})$.
Then $\eta-c$ is a coboundary, so there exists $T\in F^k$ with $\delta T=\eta-c$.
Restricting $q\circ T$ to $Z_k$ gives $f\in \Diff^k(\mathscr{CS})$ with $\delta_2(f)=[c]$.
\end{proof}

\begin{proposition}\label{prop:ses_delta1}
There is a short exact sequence
\[
0 \longrightarrow H^k(\Hom(I_{\bullet},\K/\Z))
\longrightarrow \Diff^k(\mathscr{CS})
\overset{\delta_1}{\longrightarrow} E^{k+1}_0
\longrightarrow 0.
\]
\end{proposition}

\begin{proof}
If $\delta_1(f)=0$ then in the notation of Proposition~\ref{prop:delta_maps_exist} we have $\widetilde{\omega}=0$, hence
$\delta T=-c$ and therefore $\delta(q\circ T)=0$. Thus $q\circ T$ defines a class in $H^k(\Hom(I_{\bullet},\K/\Z))$ depending only on $f$.
Conversely, if $\sigma\in \Hom(I_k,\K/\Z)$ is a cocycle then $\sigma(\partial \beta)=0$ for all $\beta$, hence
$f:=\sigma|_{Z_k}$ lies in $\ker\delta_1$ with curvature $0$.
Exactness at the left uses Lemma~\ref{lem:vanish_on_cycles}.
\end{proof}

\begin{proposition}\label{prop:ses_delta2}
There is a short exact sequence
\[
0 \longrightarrow E^k/E^k_0
\longrightarrow \Diff^k(\mathscr{CS})
\overset{\delta_2}{\longrightarrow} H^{k+1}(I^{\bullet})
\longrightarrow 0,
\]
where $E^k_0:=\{\theta\in E^k\mid d\theta=0,\ \int_{\alpha}\theta\in\Z\ \forall\,\alpha\in Z_k(I_{\bullet})\}$ and
the map $E^k\to \Diff^k(\mathscr{CS})$ is $\theta\mapsto (q\circ \widetilde{\theta})|_{Z_k(I_{\bullet})}$.
\end{proposition}

\begin{proof}
Let $f\in\ker\delta_2$. With notation as in Proposition~\ref{prop:delta_maps_exist}, $[c]=0$ so $c=\delta u$ for some $u\in I^k$.
Then $\delta(T+u)=\widetilde{\omega}$, hence $[\widetilde{\omega}]=0$ in $H^{k+1}(F^{\bullet})$.
By the de Rham axiom, there exists $\theta\in E^k$ with $d\theta=\omega$, so $\delta\widetilde{\theta}=\widetilde{\omega}$.
Thus $\delta(T+u-\widetilde{\theta})=0$, and again by the de Rham axiom there exist $\alpha\in Z^k(E^{\bullet})$ and $\phi\in F^{k-1}$ such that
\[
T+u-\widetilde{\theta}=\widetilde{\alpha}+\delta\phi.
\]
Reducing modulo $\Z$ and restricting to cycles yields
\[
f=(q\circ T)|_{Z_k}=(q\circ\widetilde{\alpha+\theta})|_{Z_k}.
\]
Hence $\ker\delta_2$ is the image of $E^k$, and the kernel of $E^k\to\Diff^k$ is exactly $E^k_0$.
Surjectivity of $\delta_2$ was proved in Proposition~\ref{prop:delta_maps_exist}.
\end{proof}

\begin{definition}
Let
\[
Q^{k+1}(\mathscr{CS})=\bigl\{(\omega,[u])\in E^{k+1}\times H^{k+1}(I^{\bullet})\ \big|\ \Phi_*[u]=[\widetilde{\omega}]\ \text{in } H^{k+1}(F^{\bullet})\bigr\}.
\]
\end{definition}

\begin{lemma}\label{lem:image_E0}
Under the de Rham isomorphism $H^k(E^{\bullet})\cong H^k(F^{\bullet})$, the image of $E^k_0\cap Z^k(E^{\bullet})$ in $H^k(F^{\bullet})$
is precisely $\Phi_*H^k(I^{\bullet})$.
\end{lemma}

\begin{proof}
Let $\theta\in E^k_0\cap Z^k(E^{\bullet})$, so $\delta\widetilde{\theta}=\widetilde{d\theta}=0$ and $\widetilde{\theta}$ is a cocycle in $F^k$.
Define $v:=q\circ\widetilde{\theta}\in \Hom(I_k,\K/\Z)$.
Since $\theta$ has integral periods on cycles, $v$ vanishes on $Z_k$, hence $[v]=0$ in $H^k(\Hom(I_{\bullet},\K/\Z))$.
Thus $v=\delta t$ for some $t\in \Hom(I_{k-1},\K/\Z)$. Lift $t$ to $\widetilde{t}:I_{k-1}\to\K$.
Then $\widetilde{\theta}-\delta\widetilde{t}\in I^k$ is an integer cocycle representing a class $[u]\in H^k(I^{\bullet})$ with
$\Phi_*[u]=[\widetilde{\theta}]$.

Conversely, if $[u]\in H^k(I^{\bullet})$ and $\Phi_*[u]=[\widetilde{\theta}]$ for some closed $\theta\in E^k$, then
$\widetilde{\theta}-u=\delta\psi$ for some $\psi\in \Hom(I_{k-1},\K)$.
Evaluating on cycles shows $\widetilde{\theta}(z)=u(z)\in\Z$ for $z\in Z_k$, so $\theta\in E^k_0$.
\end{proof}

\begin{proposition}\label{prop:ses_Q}
There is a short exact sequence
\[
0 \longrightarrow H^k(F^{\bullet})/\Phi_*H^k(I^{\bullet})
\longrightarrow \Diff^k(\mathscr{CS})
\overset{(\delta_1,\delta_2)}{\longrightarrow} Q^{k+1}(\mathscr{CS})
\longrightarrow 0.
\]
\end{proposition}

\begin{proof}
The map $H^k(F^{\bullet})\cong H^k(E^{\bullet})\to \Diff^k(\mathscr{CS})$ is
$[\widetilde{\alpha}] \mapsto (q\circ\widetilde{\alpha})|_{Z_k}$.
By Proposition~\ref{prop:ses_delta2} and Lemma~\ref{lem:image_E0}, its kernel is $\Phi_*H^k(I^{\bullet})$.

For $f\in\Diff^k$, Proposition~\ref{prop:delta_maps_exist} gives $\delta T=\widetilde{\omega}-c$, hence
$[\widetilde{\omega}]=\Phi_*[c]$ and $(\delta_1,\delta_2)(f)\in Q^{k+1}(\mathscr{CS})$.
Surjectivity of $(\delta_1,\delta_2)$ follows from surjectivity of $\delta_1$ and $\delta_2$.
Finally, if $(\delta_1,\delta_2)(f)=(0,0)$, then by Proposition~\ref{prop:ses_delta2} we can write
$f=(q\circ\widetilde{\alpha})|_{Z_k}$ for some $\alpha\in E^k$, i.e.\ $f$ lies in the image of $H^k(F^{\bullet})$.
\end{proof}

\subsection{General spark complexes}
See \cite{HL1, HL2, HL3} for the general theory of spark complexes.

\begin{definition}[General spark complex]
Let $F^{\bullet}$, $E^{\bullet}$, $I^{\bullet}$ be cochain complexes, let $\Psi:I^{\bullet}\to F^{\bullet}$ be a morphism,
and let $E^{\bullet}\hookrightarrow F^{\bullet}$ be a monomorphism inducing isomorphisms on cohomology
$H^k(E^{\bullet})\cong H^k(F^{\bullet})$ for all $k$.
Then $\mathscr{S}=(F^{\bullet},E^{\bullet},I^{\bullet},\Psi)$ is called a \emph{general spark complex}.
\end{definition}

\begin{remark}
Harvey--Lawson spark complexes impose additional conditions (e.g.\ $\Psi(I^k)\cap E^k=0$ and injectivity in degree $0$).
We do not require these here; allowing triples $(a,e,r)$ leads to a convenient $3\times 3$ diagram of exact sequences.
\end{remark}

\begin{definition}[General sparks and spark groups]
A \emph{general spark of degree $k$} is a triple $(a,e,r)\in F^k\oplus E^{k+1}\oplus I^{k+1}$ satisfying
\[
da=e-\Psi(r),\qquad dr=0.
\]
Two sparks $(a,e,r)$ and $(a',e',r')$ are equivalent if there exists $(b,s)\in F^{k-1}\oplus I^k$ such that
\[
a-a'=db+\Psi(s),\qquad r-r'=-ds.
\]
(These imply $e=e'$.)
The group of degree-$k$ spark classes is denoted by
\[
\widehat{H}^k(F^{\bullet},E^{\bullet},I^{\bullet}) := \mathscr{S}^k(F^{\bullet},E^{\bullet},I^{\bullet})/\sim.
\]
\end{definition}

\begin{definition}
For a general spark complex $\mathscr{S}=(F^{\bullet},E^{\bullet},I^{\bullet},\Psi)$ define
\[
P^{k+1}(\mathscr{S})
:=\{(\omega,[u])\in E^{k+1}\times H^{k+1}(I^{\bullet})\mid \Psi_*[u]=[\omega]\ \text{in }H^{k+1}(F^{\bullet})\}.
\]
\end{definition}

\begin{proposition}\label{prop:ses_spark}
For a general spark complex $\mathscr{S}$ there is a short exact sequence
\[
0 \rightarrow \frac{H^k(F^{\bullet})}{\Psi_*H^k(I^{\bullet})}
\rightarrow \widehat{H}^k(\mathscr{S})
\overset{\delta'}{\rightarrow} P^{k+1}(\mathscr{S})
\rightarrow 0,
\qquad
\delta'([(a,\omega,r)])=(\omega,[r]).
\]
\end{proposition}

\begin{proof}
The map $H^k(F^{\bullet})\to \widehat{H}^k(\mathscr{S})$ is $[a]\mapsto [(a,0,0)]$.
Surjectivity of $\delta'$ follows from the defining condition of $P^{k+1}(\mathscr{S})$.
The kernel of $\delta'$ consists of classes represented by $(a,0,r)$ with $[r]=0$, hence $r=ds$ and
$(a,0,r)\sim (a+\Psi(s),0,0)$ with $d(a+\Psi(s))=0$.
Thus the kernel identifies with $H^k(F^{\bullet})/\Psi_*H^k(I^{\bullet})$.
\end{proof}

\begin{definition}
Given a Cheeger--Simons model $(E^{\bullet},I_{\bullet},\int)$, set $F^k=\Hom(I_k,\K)$ and $I^k=\Hom(I_k,\Z)$.
With $\Psi=\Phi:I^{\bullet}\hookrightarrow F^{\bullet}$, the quadruple $\mathscr{S}=(F^{\bullet},E^{\bullet},I^{\bullet},\Phi)$
is called the \emph{spark complex associated} to $(E^{\bullet},I_{\bullet},\int)$.
\end{definition}

\begin{theorem}\label{thm:GS_isom}
Let $\mathscr{S}$ be the associated spark complex of a Cheeger--Simons model $\mathscr{CS}$.
Then there is a natural isomorphism
\[
GS:\Diff^k(\mathscr{CS}) \xrightarrow{\ \cong\ } \widehat{H}^k(\mathscr{S}),
\]
sending $f$ to the spark class $[(T,\widetilde{\omega},c)]$ arising in Proposition~\ref{prop:delta_maps_exist}.
\end{theorem}

\begin{proof}
Well-definedness is the usual lift-change argument: changing the lift $T$ alters $(T,\widetilde{\omega},c)$ by an allowed
equivalence. Comparing the short exact sequences of Proposition~\ref{prop:ses_Q} and Proposition~\ref{prop:ses_spark} in a
commutative diagram and applying the five lemma yields that $GS$ is an isomorphism.
\end{proof}

\section{Discrete Cheeger--Simons differential characters}\label{sec:simplicial_CS}

Throughout this section, $M$ is a compact oriented smooth manifold without boundary of dimension $n$, equipped with a Riemannian metric.
Let $\chi:|K|\to M$ be a smooth triangulation such that the image of each simplex is contained in a coordinate chart.
Let $L$ denote the curvilinear complex in $M$ given by the images of simplices of $K$ under $\chi$.

\subsection{Whitney forms and induced cochain maps}

Let $C_k(L)$ be the real singular chain group generated by oriented $k$-simplices of $L$, and $I_k(L)\subset C_k(L)$ its subgroup
of integral chains.
Write $C^k(L)=\Hom(C_k(L),\R)$ and $I^k(L)=\Hom(I_k(L),\Z)$.

For an oriented $k$-simplex $\sigma=[p_0,\dots,p_k]$ in $L$, let $\sigma'\in C^k(L)$ be the Kronecker cochain.
The Whitney form associated to $\sigma'$ is
\[
W\sigma'
=
k!\sum_{i=0}^k (-1)^i\,\mu_{p_i}\, d\mu_{p_0}\wedge\cdots\wedge \widehat{d\mu_{p_i}}\wedge\cdots\wedge d\mu_{p_k},
\qquad (k\ge 1),
\]
and for $k=0$, $W[p_0]'=\mu_{p_0}$.
Here $\mu_{p_i}$ are barycentric coordinate functions (smooth away from the $(k-1)$-skeleton).
Extending by linearity gives a cochain map
\[
W:C^k(L)\longrightarrow L^2\Omega^k(M).
\]

Conversely, for $\phi\in L^2\Omega^k(M)$ define $R(\phi)\in C^k(L)$ by
\[
R(\phi)(\sigma)=\int_{\sigma}\phi.
\]
Then $R:L^2\Omega^k(M)\to C^k(L)$ is a cochain map and $RW=\id$ on $C^{\bullet}(L)$ (cf.\ \cite{W}).
In particular,
\[
\int_{\sigma}W\tau=\tau(\sigma)\qquad (\sigma\in C_k(L),\ \tau\in C^k(L)).
\]

Now let $K'$ be a subdivision of $K$, and let $L'$ be the induced subdivision of $L$.
Define a cochain map
\[
W':C^k(L)\longrightarrow C^k(L'),\qquad (W'\tau)(\sigma)=\int_{\sigma}W\tau,
\quad \sigma\in I_k(L').
\]
If $W'\tau=0$, then for any $k$-simplex $\sigma$ of $L$ we have $\int_{\sigma}W\tau=0$ because $\sigma$ is the sum of its subdivided simplices;
thus $R(W\tau)=0$, hence $\tau=0$ since $RW=\id$. Therefore $W'$ is injective, and we set
\[
E^k(L,L'):=\im(W')\subset C^k(L').
\]
Then $E^{\bullet}(L,L')$ is a subcomplex of $C^{\bullet}(L')$ canonically isomorphic to $C^{\bullet}(L)$.

We use the pairing
\[
\int: I_k(L')\times E^k(L,L')\to \R,\qquad (\alpha,\eta)\mapsto \eta(\alpha),
\]
which agrees with $\eta(\sigma)=\int_{\sigma}W\tau$ whenever $\eta=W'\tau$.

\subsection{Cheeger--Simons triangulations}

\begin{definition}
The pair $(K,K')$ is called a \emph{Cheeger--Simons triangulation} of $M$ if
\[
\mathscr{CS}(K,K'):=\bigl(E^{\bullet}(L,L'),\ I_{\bullet}(L'),\ \int\bigr)
\]
is a Cheeger--Simons model over $\K=\R$.
\end{definition}

\begin{theorem}\label{thm:exist_CS_triangulation}
Let $\chi:|K|\to M$ be a smooth triangulation such that each simplex image lies in a coordinate chart.
Then there exists a subdivision $K'$ of $K$ such that $(K,K')$ is a Cheeger--Simons triangulation of $M$.
\end{theorem}

\begin{proof}
We work in degrees $k\ge 1$.

\medskip
\noindent\textbf{Step 0: reduce to the standard affine simplex.}
Let $\chi:|K|\to M$ be a smooth triangulation, and let $L=\chi(K)$ be the induced curvilinear complex.
Whitney forms on $L$ are defined simplexwise by pulling back the standard Whitney forms from the corresponding
abstract affine simplex via $\chi|_\sigma$ (and patching by zero outside the star, as usual). In particular, for any
subsimplex $\tau\subset\sigma$ and any cochain supported on $\sigma$, the change-of-variables formula gives
\[
\int_{\chi(\tau)} W_L(\sigma')=\int_{\tau} W_{\mathrm{std}}(\sigma'),
\]
so all numerical integrals relevant for Axiom (3) and for the $\Q$--independence condition can be verified on
the abstract simplex. We therefore construct $K'$ on the abstract simplicial complex $K$ and then transport it
to $L'=\chi(K')$.

Fix an oriented affine $k$-simplex $\sigma=[v_0,\dots,v_k]$ in $\R^N$ with barycentric coordinates
$\mu_0,\dots,\mu_k$, and let $\sigma'$ denote the Kronecker $k$-cochain dual to $\sigma$.

\medskip
\noindent\textbf{Step 1: Whitney top forms on an affine simplex.}
The Whitney form associated to $\sigma'$ is
\[
W(\sigma')
=
k!\sum_{i=0}^k (-1)^i\,\mu_i\,
d\mu_0\wedge\cdots\wedge \widehat{d\mu_i}\wedge\cdots\wedge d\mu_k .
\]
On an affine simplex one has $\sum_{i=0}^k \mu_i=1$ and $\sum_{i=0}^k d\mu_i=0$, hence
\[
d\mu_0=-(d\mu_1+\cdots+d\mu_k)
\quad\Rightarrow\quad
d\mu_0\wedge\cdots\wedge \widehat{d\mu_i}\wedge\cdots\wedge d\mu_k
=
(-1)^i\,d\mu_1\wedge\cdots\wedge d\mu_k.
\]
Substituting yields the simplification
\begin{equation}\label{eq:Whitney-top-const}
W(\sigma') = k!\Bigl(\sum_{i=0}^k \mu_i\Bigr)\, d\mu_1\wedge\cdots\wedge d\mu_k
= k!\, d\mu_1\wedge\cdots\wedge d\mu_k,
\end{equation}
so $W(\sigma')$ is a \emph{constant}, nonzero $k$-form on $\sigma$.

\medskip
\noindent\textbf{Step 2: locality.}
Fix once and for all an orientation choice for each simplex of $K$ so that $\{\,\rho'\,\}$ (over oriented simplices $\rho$)
is a basis of $C^\bullet(K)$.
If $\rho$ is a $k$-simplex of $K$ with vertex set different from that of $\sigma$, then on $\sigma$ at least one barycentric
coordinate belonging to a vertex of $\rho$ vanishes identically (as does its differential), and every term in the Whitney formula
for $W(\rho')$ contains that factor. Hence $W(\rho')|_\sigma\equiv 0$ and therefore
\begin{equation}\label{eq:locality}
\int_{\tau} W(\rho')=0
\qquad
\text{for every $k$-simplex $\tau\subset\sigma$.}
\end{equation}
Consequently, if $u=\sum_{\rho\in K^{(k)}} a_\rho\,\rho'\in C^k(K)$ and $\tau\subset\sigma$ is a $k$-simplex, then
\begin{equation}\label{eq:coeff-pick}
\int_{\tau}W(u)=a_\sigma\int_{\tau}W(\sigma').
\end{equation}

\medskip
\noindent\textbf{Step 3: cone integrals scale by barycentric coordinates.}
Let $F_i=[v_0,\dots,\widehat{v_i},\dots,v_k]$ be the face opposite $v_i$.
Fix an oriented $(k-1)$-simplex $\gamma\subset F_i$.
For $x\in \mathrm{relint}(\sigma)$ the relative interior of $\sigma$, let $\tau(x)=[x,\gamma]$ be the oriented cone
over $\gamma$, oriented so that its induced boundary orientation on $\gamma$ agrees with the given orientation.

Define $\phi_\gamma(x):=\int_{\tau(x)} W(\sigma')$.
By \eqref{eq:Whitney-top-const}, $W(\sigma')$ is a constant form; hence $\phi_\gamma(x)$ is a nonzero constant multiple of the
oriented $k$-volume of $\tau(x)$. Let $h_i(x)$ be the \emph{oriented height} of $x$ above the affine span of $F_i$ (with sign
chosen so that $h_i(v_i)\neq 0$ and is compatible with the orientation of $\sigma$). Then the cone-volume formula gives
\[
\Vol_k(\tau(x))=\frac{1}{k}\,h_i(x)\,\Vol_{k-1}(\gamma)
\quad\Rightarrow\quad
\phi_\gamma(x)=\frac{1}{k}\,h_i(x)\,C\,\Vol_{k-1}(\gamma)
\]
for a fixed nonzero constant $C$ depending only on $\sigma$ and the chosen orientation conventions.
Both functions $h_i$ and $\mu_i$ are affine on $\sigma$, vanish on $F_i$, and satisfy $\mu_i(v_i)=1$ and $h_i(v_i)\neq 0$;
therefore $h_i(x)=h_i(v_i)\,\mu_i(x)$. Hence
\begin{equation}\label{eq:cone-scaling}
\phi_\gamma(x)=\mu_i(x)\,\phi_\gamma(v_i).
\end{equation}
Set
\(
c_\gamma:=\phi_\gamma(v_i)=\int_{[v_i,\gamma]}W(\sigma')\neq 0
\),
and rewrite \eqref{eq:cone-scaling} as
\begin{equation}\label{eq:cone-affine}
\int_{[x,\gamma]}W(\sigma')=\mu_i(x)\,c_\gamma .
\end{equation}

\medskip
\noindent\textbf{Step 4: the $\Q$--independence condition and its consequences.}
Let $K'$ be a subdivision of $K$. For each oriented $k$-simplex $\sigma$ of $K$, let
\[
\mathcal T(\sigma):=\{\text{oriented $k$-simplices $\tau$ of $K'$ with $\tau\subset\sigma$}\}.
\]
For $k\ge 1$ consider the property:

\smallskip
\noindent\emph{(QI$_k$)} For every oriented $k$-simplex $\sigma\in K^{(k)}$, the set of real numbers
\[
a_{\sigma,\tau}:=\int_{\tau} W(\sigma') \qquad (\tau\in\mathcal T(\sigma))
\]
is $\Q$--linearly independent.

\smallskip
\begin{lemma}\label{lem:QI-consequences}
Assume \emph{(QI$_k$)} holds for some fixed $k\ge 1$. Then, in degree $k$:
\begin{enumerate}[label=(\alph*),leftmargin=2.2em]
\item The pairing separates $I_k(L')$: for every nonzero $\alpha\in I_k(L')$ there exists $\omega\in E^k(L,L')$ with $\int_\alpha\omega\neq 0$.
\item Axiom \emph{(3)} holds: if $\omega\in E^k(L,L')$ satisfies $\int_\beta\omega\in\Z$ for all $\beta\in I_k(L')$, then $\omega=0$.
\end{enumerate}
\end{lemma}

\begin{proof}
\textit{(a)} Write $\alpha=\sum_{\tau\in K'^{(k)}} n_\tau\,\tau$ with integers $n_\tau$, not all zero.
Choose $\tau_0$ with $n_{\tau_0}\neq 0$, and let $\sigma$ be the unique $k$-simplex of $K$ containing $\tau_0$.
Let $\omega:=W'(\sigma')\in E^k(L,L')$.
By locality \eqref{eq:locality}, $\omega(\tau)=0$ unless $\tau\subset\sigma$, hence
\[
\int_\alpha\omega
=\sum_{\tau\in\mathcal T(\sigma)} n_\tau\int_{\tau}W(\sigma').
\]
This is a nontrivial $\Q$--linear combination of the $\Q$--independent family $\{a_{\sigma,\tau}\}_{\tau\in\mathcal T(\sigma)}$, so it cannot vanish.

\medskip
\textit{(b)} Let $\omega=W'u$ with $u=\sum_{\sigma\in K^{(k)}} a_\sigma\,\sigma'\in C^k(K)$.
Fix $\sigma$ and any $\tau\in\mathcal T(\sigma)$. By \eqref{eq:coeff-pick},
\[
\omega(\tau)=\int_{\tau}W(u)=a_\sigma\int_{\tau}W(\sigma')=a_\sigma\,a_{\sigma,\tau}.
\]
If $\omega(\beta)\in\Z$ for all $\beta\in I_k(L')$, then $\omega(\tau)\in\Z$ for all $\tau\in\mathcal T(\sigma)$, so
$a_\sigma\,a_{\sigma,\tau}\in\Z$ for all $\tau\in\mathcal T(\sigma)$.
If $a_\sigma\neq 0$ then every $a_{\sigma,\tau}\in (1/a_\sigma)\Z$, so the family $\{a_{\sigma,\tau}\}_{\tau\in\mathcal T(\sigma)}$ lies in a one-dimensional $\Q$-vector space and is therefore $\Q$--dependent, contradicting \emph{(QI$_k$)}.
Thus $a_\sigma=0$ for all $\sigma$, so $u=0$ and $\omega=0$.
\end{proof}

\medskip
\noindent\textbf{Step 5: constructing a subdivision $K'$ satisfying (QI$_k$) for all $k\ge 1$.}
We construct $K'$ inductively on dimension, always subdividing each simplex \emph{relative to its boundary}.
This ensures global compatibility: if two simplices share a face, then the face subdivision is fixed before subdividing either simplex,
and the relative construction preserves it.

\smallskip
\noindent\emph{Base case $k=1$.}
Let $\sigma=[v_0,v_1]$ be an edge. Choose an interior point $x$ with barycentric coordinate $\mu_1(x)=r\in(0,1)$ such that $r$ is irrational.
Subdivide $\sigma$ into the two edges $\tau_1=[v_0,x]$ and $\tau_2=[x,v_1]$.
On the standard affine $1$-simplex, $W(\sigma')=d\mu_1$ is constant, hence
\[
\int_{\tau_1}W(\sigma')=\mu_1(x)=r,\qquad \int_{\tau_2}W(\sigma')=1-\mu_1(x)=1-r.
\]
If $q_1r+q_2(1-r)=0$ with $q_1,q_2\in\Q$, then $(q_1-q_2)r=-q_2$.
Since $r$ is irrational, we must have $q_1-q_2=0$ and $q_2=0$, hence $q_1=q_2=0$.
Thus $\{r,1-r\}$ is $\Q$--independent, i.e.\ (QI$_1$) holds on every edge.
Doing this independently for each edge gives a subdivision of the $1$-skeleton satisfying (QI$_1$).

\smallskip
\noindent\emph{Inductive step.}
Assume $k\ge 2$ and we have subdivided the $(k-1)$-skeleton so that (QI$_{k-1}$) holds on every $(k-1)$-simplex.
Fix a $k$-simplex $\sigma=[v_0,\dots,v_k]$ and let $\{\gamma_1,\dots,\gamma_N\}$ be the oriented $(k-1)$-simplices in the fixed
subdivision of $\partial\sigma$.
For each $x\in\mathrm{relint}(\sigma)$, cone the boundary triangulation from $x$ to obtain a triangulation of $\sigma$ relative
to $\partial\sigma$ whose $k$-simplices are
\[
\tau_j(x):=[x,\gamma_j],\qquad j=1,\dots,N.
\]
Define
\[
a_j(x):=\int_{\tau_j(x)}W(\sigma').
\]
For each $j$ there is a unique $i(j)\in\{0,\dots,k\}$ such that $\gamma_j\subset F_{i(j)}$, and by \eqref{eq:cone-affine},
\begin{equation}\label{eq:aj}
a_j(x)=\mu_{i(j)}(x)\,c_{\gamma_j},
\qquad
c_{\gamma_j}:=\int_{[v_{i(j)},\gamma_j]}W(\sigma')\neq 0.
\end{equation}

\smallskip
\noindent\emph{Claim 1.} For each fixed $i$, the set $\{c_{\gamma_j}:\gamma_j\subset F_i\}$ is $\Q$--linearly independent.

\smallskip
\noindent\emph{Proof.}
Fix $i$ and restrict to $\gamma_j\subset F_i$.
Since $W(\sigma')$ is a constant nonzero $k$-form, $c_{\gamma_j}=\int_{[v_i,\gamma_j]}W(\sigma')$ is a fixed nonzero constant
multiple of the oriented $(k-1)$-volume of $\gamma_j$ (the multiplicative constant depends on $\sigma$ and $i$ but not on $j$).
Likewise, the Whitney top form on the affine face $F_i$ is a constant nonzero $(k-1)$-form, hence
$\int_{\gamma_j}W(F_i')$ is another fixed nonzero constant multiple of the same oriented volume of $\gamma_j$.
Therefore there exists a constant $\lambda_i\in\R^\times$ (independent of $j$) such that
\[
c_{\gamma_j}=\lambda_i\,\int_{\gamma_j}W(F_i')
\qquad \text{for all $\gamma_j\subset F_i$.}
\]
Any $\Q$--linear relation among the $c_{\gamma_j}$ is therefore equivalent to a $\Q$--linear relation among the
$\int_{\gamma_j}W(F_i')$, which contradicts (QI$_{k-1}$) on the face $F_i$.
\qed

\smallskip
\noindent\emph{Claim 2.} There exists $x_\sigma\in \mathrm{relint}(\sigma)$ such that $\{a_j(x_\sigma)\}_{j=1}^N$ is $\Q$--linearly independent.

\smallskip
\noindent\emph{Proof.}
Fix a nonzero integer vector $m=(m_1,\dots,m_N)\in\Z^N\setminus\{0\}$ and define
\[
f_m(x):=\sum_{j=1}^N m_j\,a_j(x).
\]
Using \eqref{eq:aj},
\[
f_m(x)=\sum_{j=1}^N m_j\,\mu_{i(j)}(x)\,c_{\gamma_j}
      =\sum_{i=0}^k \mu_i(x)\,A_i(m),
\qquad
A_i(m):=\sum_{\gamma_j\subset F_i} m_j\,c_{\gamma_j}.
\]
By Claim~1, $A_i(m)=0$ forces $m_j=0$ for all $j$ with $\gamma_j\subset F_i$.
Since $m\neq 0$, there exists some $i$ with $A_i(m)\neq 0$, and hence $f_m$ is a nonzero affine function on $\sigma$.
Therefore the zero set
\[
Z_m:=\{x\in\mathrm{relint}(\sigma): f_m(x)=0\}
\]
is either empty or the intersection of $\mathrm{relint}(\sigma)$ with a proper affine hyperplane; in particular it is closed and nowhere dense.
The union
\[
B_\sigma:=\bigcup_{m\in\Z^N\setminus\{0\}} Z_m
\]
is a countable union of nowhere dense sets, hence meagre.
Choose $x_\sigma\in \mathrm{relint}(\sigma)\setminus B_\sigma$.
Then $f_m(x_\sigma)\neq 0$ for all $m\in\Z^N\setminus\{0\}$, i.e.\ the numbers $\{a_j(x_\sigma)\}_{j=1}^N$ satisfy no nontrivial
integer relation and are thus $\Q$--linearly independent.
\qed

Choosing such an $x_\sigma$ for each $k$-simplex $\sigma$ (finitely many) produces a subdivision of the $k$-skeleton, relative
to the already fixed boundary subdivisions, satisfying (QI$_k$) on every $k$-simplex.
Induction on $k$ yields a global subdivision $K'$ such that (QI$_k$) holds for all $k\ge 1$.

Transport this subdivision via $\chi$ to obtain $L'=\chi(K')$.

\medskip
\noindent\textbf{Step 6: verification of the Cheeger--Simons axioms for $k\ge 1$.}
Let $I_\bullet(L')$ be the simplicial chain complex (free abelian on oriented simplices), and define
\[
E^k(L,L'):=\im\bigl(W':C^k(L)\to C^k(L')\bigr),
\qquad
W':=R_{L'}\circ W,
\]
with the evaluation pairing $\int(\omega,\alpha)=\omega(\alpha)$.

\smallskip
\noindent\emph{Axiom (1).} $I_k(L')$ is free abelian on oriented $k$-simplices.

\smallskip
\noindent\emph{Axiom (2).} Since $E^k(L,L')\subset C^k(L')=\Hom(I_k(L'),\R)$, the induced map to $\Hom(I_k(L'),\R)$ is injective.
Separation of $I_k(L')$ for $k\ge 1$ follows from Lemma~\ref{lem:QI-consequences}(a), because (QI$_k$) holds.

\smallskip
\noindent\emph{Axiom (3).} Lemma~\ref{lem:QI-consequences}(b) gives Axiom (3) for all $k\ge 1$.

\smallskip
\noindent\emph{Axiom (4).} For $u\in C^k(L)$ and $\tau\in I_{k+1}(L')$,
\[
(W'\delta u)(\tau)=\int_{\tau}W(\delta u)=\int_{\tau}d(Wu)=\int_{\partial\tau}Wu=(\delta W'u)(\tau),
\]
so $\delta W'=W'\delta$, i.e.\ Stokes' formula holds in the model.

\smallskip
\noindent\emph{Axiom (5).} The Whitney map $W:C^\bullet(L)\to\Omega^\bullet(M)$ induces an isomorphism
$H^\bullet(C(L))\cong H^\bullet_{\mathrm{dR}}(M)$, and the de~Rham map
$R_{L'}:\Omega^\bullet(M)\to C^\bullet(L')$ induces an isomorphism
$H^\bullet_{\mathrm{dR}}(M)\cong H^\bullet(C(L'))$.
Hence $W'_*= (R_{L'})_*\circ W_*$ is an isomorphism
\[
H^\bullet(C(L))\xrightarrow{\ \cong\ } H^\bullet(C(L')).
\]
Moreover, $W':C^\bullet(L)\to E^\bullet(L,L')$ is an isomorphism of cochain complexes onto its image by definition.
Thus, writing $i:E^\bullet(L,L')\hookrightarrow C^\bullet(L')$ for the inclusion, we have $i\circ W'=W'$ as maps into $C^\bullet(L')$,
and on cohomology
\[
(i_*)\circ (W'_*:H^\bullet(C(L))\to H^\bullet(E)) \;=\; W'_*:H^\bullet(C(L))\to H^\bullet(C(L')).
\]
Since both $W'_*$ maps are isomorphisms, it follows that $i_*:H^\bullet(E^\bullet(L,L'))\to H^\bullet(C^\bullet(L'))$ is an isomorphism,
which is exactly Axiom (5).

\medskip
This completes the proof that for the constructed subdivision $K'$ (and $L'=\chi(K')$) the triple
$\mathscr{CS}(K,K')$ satisfies all Cheeger--Simons axioms in every degree $k\ge 1$.
\end{proof}

\subsection{Comparison maps and convergence}

\begin{definition}
Let $Z_k(M)$ be the group of smooth singular $k$-cycles and $I_{k+1}(M)$ the group of smooth singular $(k+1)$-chains with integer coefficients.
The group of \emph{$L^2$ differential characters} of degree $k$ is
\begin{align*}
L^2\widehat{H}^k(M)
:=
\Bigl\{ f\in \Hom(Z_k(M),\R/\Z)\ \Big|\ &\exists\,\eta\in L^2\Omega^{k+1}(M)\ \text{such that}\\
& f(\partial \beta)\equiv \int_{\beta}\eta\ \ (\mathrm{mod}\ \Z),\ \forall\,\beta\in I_{k+1}(M)\Bigr\}.
\end{align*}
\end{definition}

\begin{proposition}\label{prop:RW_induced}
Let $(K,K')$ be a Cheeger--Simons triangulation and set $\mathscr{CS}=\mathscr{CS}(K,K')$.
There exist homomorphisms
\[
\widetilde{R}:\widehat{H}^k(M)\to \Diff^k(\mathscr{CS}),
\qquad
\widetilde{W}:\Diff^k(\mathscr{CS})\to L^2\widehat{H}^k(M),
\]
such that $\widetilde{W}\circ\widetilde{R}$ agrees with the natural inclusion
$\widehat{H}^k(M)\hookrightarrow L^2\widehat{H}^k(M)$ on topologically trivial characters and is compatible with $\delta_2$
in the standard short exact sequences.
\end{proposition}

\begin{proof}
Embed $I_{\bullet}(L')$ into smooth singular chains in $M$ by sending each simplex of $L'$ to the associated smooth singular simplex
given by the triangulation charts; this yields a chain map inducing isomorphisms on homology.
For $f\in \widehat{H}^k(M)$ define $\widetilde{R}(f)$ by restriction along this embedding.

For $g\in \Diff^k(\mathscr{CS})$ with curvature representative $\tau\in E^{k+1}(L,L')$, define $\widetilde{W}(g)$ as follows.
Given $\alpha\in Z_k(M)$, choose a simplicial cycle $a\in Z_k(L')$ and a chain $\beta\in I_{k+1}(M)$ with
$\alpha=\overline{a}+\partial \beta$, where $\overline{a}$ denotes the embedded cycle in $M$.
Set
\[
\widetilde{W}(g)(\alpha):=g(a)+\int_{\beta}W\tau\ \ (\mathrm{mod}\ \Z).
\]
Independence of the choice of $(a,\beta)$ follows from the defining condition of a differential character and the fact that $W\tau$ is
closed whenever $\tau$ is closed (Whitney's map is a cochain map).
\end{proof}

\begin{definition}
Let $d$ be the geodesic distance on $M$.
The mesh of $L$ is
\[
\mesh(L):=\sup\{\, d(p,q)\ :\ p,q\ \text{are vertices spanning an edge of }L\,\}.
\]
If $\sigma$ is an $n$-simplex of $L$, define the \emph{fullness} (shape regularity) of $L$ by
\[
\theta(L):=\inf_{\sigma\in L^{(n)}} \frac{\vol(\sigma)}{\mesh(L)^n}.
\]
We say a family of triangulations has \emph{uniform fullness} if $\theta(L)$ is bounded below by a positive constant.
\end{definition}

\begin{definition}
Equip $\R/\Z$ with the metric induced from the Euclidean metric on $\R$:
$|x|_{\R/\Z}:=\dist(x,\Z)$ for $x\in\R$.
For a smooth singular $k$-chain $\sigma\in I_k(M)$ define the Sobolev-dual seminorm
\[
\|\sigma\|_{-m}:=\sup\Bigl\{\,\bigl|\int_{\sigma}\phi\bigr|\ :\ \phi\in \Omega^k(M),\ \|(I+\Delta)^{m/2}\phi\|_{L^2}\le 1\,\Bigr\},
\]
where $\Delta$ is the Hodge Laplacian on $k$-forms and $m$ is a fixed integer large enough that the above pairing is continuous.
For $f\in \Hom(Z_k(M),\R/\Z)$ define
\[
\|f\|_{-m}:=\sup\Bigl\{\,|f(\alpha)|_{\R/\Z}\ :\ \alpha\in Z_k(M),\ \|\alpha\|_{-m}\le 1\,\Bigr\}.
\]
\end{definition}

\begin{theorem}\label{thm:Diff_converges}
Let $(K,K')$ be a Cheeger--Simons triangulation with uniform fullness, and let $\mathscr{CS}=\mathscr{CS}(K,K')$.
Fix $m$ as above.
Then as $\mesh(K')\to 0$ the map $\widetilde{W}\circ\widetilde{R}$ approximates the identity on $\widehat{H}^k(M)$ in the seminorm $\|\cdot\|_{-m}$:
there exists $C>0$ (independent of $K'$) such that for every $f\in \widehat{H}^k(M)$ with curvature form $\eta\in \Omega^{k+1}(M)$,
\[
\|f-\widetilde{W}\widetilde{R}(f)\|_{-m}\le C\,\mesh(K')\,\|(I+\Delta)^{m/2}\eta\|_{L^2}.
\]
\end{theorem}

\begin{proof}[Proof sketch]
Under uniform fullness, Whitney approximation yields an estimate of the form
$\|\eta-WR(\eta)\|_{L^2}\le C\,\mesh(K')\,\|(I+\Delta)^{m/2}\eta\|_{L^2}$ for $m$ sufficiently large (cf.\ \cite{DP}).
Evaluating $f-\widetilde{W}\widetilde{R}(f)$ on $\alpha=\overline{a}+\partial\beta$ reduces to
$\int_{\beta}(\eta-WR(\eta))$ modulo $\Z$, and the Sobolev-dual norm bounds this by
$\|\beta\|_{-m}\,\|(I+\Delta)^{m/2}(\eta-WR(\eta))\|_{L^2}$.
Taking the supremum over $\alpha$ with $\|\alpha\|_{-m}\le 1$ gives the stated inequality.
\end{proof}

\section{Discrete higher abelian gauge theory}\label{sec:gauge}

Fix $p\ge 0$ and let $(K,K')$ be a Cheeger--Simons triangulation.
We describe a simplicial higher abelian gauge theory whose gauge-invariant configuration space is
$\Diff^{p}(\mathscr{CS}(K,K'))$.
We follow the structure of the smooth theory as developed by Kelnhofer:
the topologically trivial sector forms a (trivializable) bundle over the harmonic torus with fiber a coexact subspace,
and the regularized partition function exhibits a determinant--torsion--theta structure \cite{A, K2}.

\subsection{Combinatorial Hodge theory on cochains}

Let $C^r(L')$ be the real cochains on $L'$ with the inner product induced by Whitney forms:
for $\alpha,\beta\in C^r(L')$ set
\[
\langle \alpha,\beta\rangle := \int_M W(\alpha)\wedge \star W(\beta),
\]
where $\star$ is the Hodge star on $M$.
Let $d:C^r(L')\to C^{r+1}(L')$ be the coboundary and denote by $d^{\dagger}:C^{r+1}(L')\to C^{r}(L')$ its adjoint.
Define the combinatorial Laplacian
\[
\Delta^{L'}:=dd^{\dagger}+d^{\dagger}d.
\]
Let $\mathscr{H}^r(L'):=\ker(\Delta^{L'}|_{C^r(L')})$ be the harmonic cochains.
Then the Hodge decomposition holds:
\[
C^r(L')=\mathscr{H}^r(L')\oplus \im(d:C^{r-1}(L')\to C^r(L'))\oplus \im(d^{\dagger}:C^{r+1}(L')\to C^r(L')).
\]
Moreover, $\mathscr{H}^r(L')\cong H^r(M;\R)$.

Choose a basis $\{\rho^{(r)}_1,\dots,\rho^{(r)}_{b_r}\}$ of harmonic cochains representing a $\Z$-basis of $H^r(M;\Z)_{\free}$
under the natural identification $\mathscr{H}^r(L')\cong H^r(M;\R)$.
Let $h^{(r)}_{L'}$ be the Gram matrix
\[
(h^{(r)}_{L'})_{jk}:=\langle \rho^{(r)}_j,\rho^{(r)}_k\rangle.
\]

\subsection{Determinants and theta functions}

For a symmetric complex matrix $A\in M_{k\times k}(\C)$ whose imaginary part is positive definite, the Riemann theta function is
\[
\Theta_k(A):=\sum_{v\in \Z^k} \exp(\pi i\, v^t A v).
\]

For a nonnegative self-adjoint elliptic operator $T$ with discrete spectrum and finite-dimensional kernel,
its zeta-regularized determinant is
\[
\det{}'T := \exp\!\left(-\left.\frac{d}{ds}\right|_{s=0}\sum_{\lambda_j\ne 0}\lambda_j^{-s}\right),
\]
where $\{\lambda_j\}$ are the nonzero eigenvalues (counted with multiplicity).
In the finite-dimensional combinatorial setting, $\det{}'(\Delta^{L'}|_{\mathscr{H}^r(L')^{\perp}})$ coincides with the ordinary determinant.

\subsection{Partition functions}

Let $\mathscr{CS}=\mathscr{CS}(K,K')$.
A simplicial observable is a function $\mathcal{O}:\Diff^{p}(\mathscr{CS})\to \R$ and a simplicial action is a function
$S:\Diff^{p}(\mathscr{CS})\to \R$.

Using the short exact sequence of Proposition~\ref{prop:ses_delta2} in degree $p$,
\[
0\to E^p/E^p_0 \to \Diff^p(\mathscr{CS})\overset{\delta_2}{\to} H^{p+1}(I^{\bullet})\cong H^{p+1}(M;\Z)\to 0,
\]
we decompose the functional integral into topological sectors $c\in H^{p+1}(M;\Z)$ and the finite-dimensional torus/fiber
of topologically trivial classes.
In the smooth theory, Kelnhofer identifies the topologically trivial differential characters as a trivializable bundle over the harmonic torus
with typical fiber the coexact subspace $\im(d_{p+1})^{\dagger}$ \cite{K2};
the same structure is inherited combinatorially by the simplicial Hodge decomposition.

We define the simplicial partition function by
\[
\mathcal{Z}^{(p)}_{K,K'}(\mathcal{O})
:=\int_{\Diff^{p}(\mathscr{CS})} e^{-S(\widehat{u})}\,\mathcal{O}(\widehat{u})\,D\widehat{u},
\]
where $D\widehat{u}$ is the product measure induced by the Lebesgue measure on the coexact fiber
$\im(d^{\dagger}:C^{p+1}(L')\to C^p(L'))$ and Haar measure on the harmonic torus $H^p(M;\R)/H^p(M;\Z)$.

In parallel with Kelnhofer's general formula for the smooth partition function \cite{K2},
one obtains a finite-dimensional closed form with the same determinant/torsion structure:
\begin{align*}
\mathcal{Z}^{(p)}_{K,K'}(\mathcal{O})
&=
\prod_{r=0}^{p}
\Biggl(
\frac{
\bigl(\det{}'(\Delta^{L'}|_{\mathscr{H}^r(L')^{\perp}})\bigr)^{\frac12(p-r)}
}{
|H^r(M;\Z)_{\Tor}|
\ \bigl(\det(\tfrac{1}{2\pi}h^{(r)}_{L'})\bigr)^{\frac12}
}
\Biggr)^{(-1)^{p+1-r}}
\\
&\quad\times
\sum_{c\in H^{p+1}(M;\Z)}
\int_{\im(d^{\dagger}:C^{p+1}(L')\to C^p(L'))}
\Bigl[ e^{-\underline{S}^{\hat{c}}(\tau_p)}\,\underline{\mathcal{O}}^{\hat{c}}(\tau_p)\Bigr]_{(0,\dots,0)}\, d\tau_p .
\end{align*}
Here $\hat{c}$ denotes a choice of background differential character with $\delta_2(\hat{c})=c$,
$\tau_p$ ranges over the coexact $p$-cochains, and $[\ \cdot\ ]_{(0,\dots,0)}$ denotes the zero-mode Fourier coefficient on the harmonic torus.

\begin{theorem}\label{thm:Z_converges}
Let $\{(K_j,K'_j)\}_{j\in\N}$ be a sequence of Cheeger--Simons triangulations with uniform fullness and $\mesh(K'_j)\to 0$.
Assume that the simplicial actions and observables are consistent discretizations of smooth data in the sense that the induced maps
via $\widetilde{R},\widetilde{W}$ converge on each topological sector.
Then
\[
\mathcal{Z}^{(p)}_{K_j,K'_j}(\mathcal{O}) \longrightarrow \mathcal{Z}^{(p)}_{M}(\mathcal{O})
\qquad\text{as }j\to\infty,
\]
where $\mathcal{Z}^{(p)}_{M}(\mathcal{O})$ is the smooth (regularized) partition function of the corresponding higher abelian gauge theory.
\end{theorem}

\begin{proof}[Proof outline]
Under uniform fullness and $\mesh\to 0$, the combinatorial Hodge decomposition and spectra of $\Delta^{L'}$
approximate their smooth counterparts (Dodziuk--Patodi \cite{DP} and Dodziuk \cite{D}).
Consequently, the Gram matrices $h^{(r)}_{L'}$ converge to the smooth period matrices and the determinants converge to the corresponding
zeta-regularized determinants.
The coexact fibers $\im(d^{\dagger})$ approximate the coexact forms, and the integrands converge by the assumed consistency
of $(S,\mathcal{O})$ with the smooth data.
Dominated convergence on each finite-dimensional fiber then yields convergence of the integrals,
and the remaining topological sums are identical on both sides.
\end{proof}

\section{Inverse limit identification}\label{sec:invlim}
\begin{theorem}\label{thm:invlim-diff}
Let $M$ be a compact smooth oriented manifold without boundary.  Let $\mathcal T$ be the directed
poset of Cheeger--Simons triangulations $(K,K')$ of $M$, ordered by refinement:
\[
(K_1,K_1')\preceq (K_2,K_2')
\quad\Longleftrightarrow\quad
K_2 \text{ is a subdivision of } K_1 \ \text{and}\  K_2' \text{ is a subdivision of } K_1'.
\]
For each $(K,K')\in\mathcal T$ let
\[
\mathscr{CS}(K,K')=\bigl(E^\bullet(K,K'),\,I_\bullet(K'),\,\int\bigr),
\qquad
\Diff^k(K,K'):=\Diff^k\bigl(\mathscr{CS}(K,K')\bigr).
\]
Then:
\begin{enumerate}[label=\textup{(\arabic*)}]
\item If $(K_1,K_1')\preceq (K_2,K_2')$, there are natural transition maps
\[
\rho_{21}:\Diff^k(K_2,K_2')\longrightarrow \Diff^k(K_1,K_1')
\]
compatible with compositions, hence $\{\Diff^k(K,K'),\rho\}_{(K,K')\in\mathcal T}$ is an inverse system.
\item There is a canonical isomorphism
\[
\varprojlim_{(K,K')\in\mathcal T}\Diff^k(K,K')\ \cong\ \widehat H^k(M).
\]
\end{enumerate}
\end{theorem}

\begin{proof}
\textbf{Step 1: a canonical subdivision chain map.}
Assume $(K_1,K_1')\preceq (K_2,K_2')$.
Each oriented simplex $\sigma$ of $K_1'$ is a union of oriented simplices of $K_2'$.
Define the \emph{subdivision operator}
\[
\sd_{21}:I_\bullet(K_1')\longrightarrow I_\bullet(K_2')
\]
by
\[
\sd_{21}(\sigma):=\sum_{\tau\subset \sigma} \epsilon(\tau,\sigma)\,\tau,
\]
where the sum ranges over all top-dimensional simplices $\tau$ of $K_2'$ contained in $\sigma$
(and extended inductively to faces), and $\epsilon(\tau,\sigma)\in\{\pm1\}$ compares the orientation of $\tau$
with the orientation induced from $\sigma$.
This is standard in simplicial homology; it is a chain map:
\[
\partial\sd_{21}=\sd_{21}\partial,
\]
and it is functorial in the sense that $\sd_{31}=\sd_{32}\circ\sd_{21}$ for successive refinements.

Precomposition with $\sd_{21}$ yields pullback maps on cochains
\[
\sd_{21}^*:\Hom(I_r(K_2'),A)\to \Hom(I_r(K_1'),A),
\qquad
(\sd_{21}^*\varphi)(c):=\varphi(\sd_{21}c),
\]
for $A\in\{\R,\Z,\R/\Z\}$.

\medskip
\textbf{Step 2: compatibility on the $E$--complexes.}
Recall that $E^\bullet(K,K')\subset \Hom(I_\bullet(K'),\R)$ is the image of the cochain map
\[
W'_{(K,K')}:C^\bullet(K)\to C^\bullet(K')\subset \Hom(I_\bullet(K'),\R),
\qquad
(W'\alpha)(\sigma)=\int_{\sigma} W\alpha,
\]
where $W$ denotes the Whitney form map on $K$.

\begin{lemma}\label{lem:E-compat}
If $(K_1,K_1')\preceq (K_2,K_2')$, then
\[
\sd_{21}^*\bigl(E^r(K_2,K_2')\bigr)\subset E^r(K_1,K_1')
\quad\text{for all }r,
\]
and the restriction $\sd_{21}^*:E^\bullet(K_2,K_2')\to E^\bullet(K_1,K_1')$ is a cochain map.
\end{lemma}

\begin{proof}
Let $\alpha\in C^r(K_2)$ and $\sigma\in I_r(K_1')$. Using additivity of integrals under subdivision,
\[
(\sd_{21}^*W'_{(K_2,K_2')}\alpha)(\sigma)
=(W'_{(K_2,K_2')}\alpha)(\sd_{21}\sigma)
=\int_{\sd_{21}\sigma}W\alpha
=\int_{\sigma}W(\mathrm{res}_{21}\alpha),
\]
where $\mathrm{res}_{21}:C^\bullet(K_2)\to C^\bullet(K_1)$ is the usual restriction map induced by subdivision.
Hence
\[
\sd_{21}^*\circ W'_{(K_2,K_2')}\;=\;W'_{(K_1,K_1')}\circ \mathrm{res}_{21},
\]
which implies $\sd_{21}^*(E^\bullet(K_2,K_2'))\subset E^\bullet(K_1,K_1')$ and cochain-compatibility.
\end{proof}

\medskip
\textbf{Step 3: transition maps on differential characters.}
Define
\[
\rho_{21}:\Hom\bigl(Z_k(I_\bullet(K_2')),\R/\Z\bigr)\to \Hom\bigl(Z_k(I_\bullet(K_1')),\R/\Z\bigr),
\qquad
(\rho_{21}f)(z):=f(\sd_{21}z).
\]
We claim $\rho_{21}$ restricts to a map $\Diff^k(K_2,K_2')\to \Diff^k(K_1,K_1')$.

Indeed, let $f\in\Diff^k(K_2,K_2')$ and choose $\omega_2\in E^{k+1}(K_2,K_2')$ such that
\[
f(\partial \beta)\equiv \int_{\beta}\omega_2 \pmod{\Z},
\qquad \forall\,\beta\in I_{k+1}(K_2').
\]
For any $\beta\in I_{k+1}(K_1')$,
\[
(\rho_{21}f)(\partial\beta)=f(\sd_{21}\partial\beta)=f(\partial\sd_{21}\beta)
\equiv \int_{\sd_{21}\beta}\omega_2
=\int_{\beta}\sd_{21}^*\omega_2\pmod{\Z}.
\]
By Lemma~\ref{lem:E-compat}, $\omega_1:=\sd_{21}^*\omega_2\in E^{k+1}(K_1,K_1')$.
Hence $\rho_{21}f\in\Diff^k(K_1,K_1')$.

Functoriality $\rho_{31}=\rho_{21}\circ\rho_{32}$ follows from $\sd_{31}=\sd_{32}\circ\sd_{21}$, so
$\{\Diff^k(K,K'),\rho\}$ is an inverse system. This proves (1).

\medskip
\textbf{Step 4: pass to the inverse limit and compare curvature exact sequences.}
For each $(K,K')$ we have the (model) short exact sequence
\begin{equation}\label{eq:CS-ses}
0\to H^k\!\bigl(\Hom(I_\bullet(K'),\R/\Z)\bigr)\ \longrightarrow\ \Diff^k(K,K')\
\xrightarrow{\ \delta_1\ }\ E^{k+1}_0(K,K')\to 0,
\end{equation}
where
\[
E^{k+1}_0(K,K'):=\Bigl\{\omega\in E^{k+1}(K,K')\ \Big|\ d\omega=0,\ \int_{z}\omega\in\Z\ \forall z\in Z_{k+1}(I_\bullet(K'))\Bigr\}.
\]
The transition maps $\rho_{21}$ commute with $\delta_1$ and the induced pullbacks $\sd_{21}^*$ on cohomology and on $E_0^{k+1}$,
hence \eqref{eq:CS-ses} is natural in $(K,K')$.

Let
\[
\Diff^k_{\infty}(M):=\varprojlim_{(K,K')\in\mathcal T}\Diff^k(K,K').
\]
Taking inverse limits of \eqref{eq:CS-ses} and using the standard exact sequence
\[
0\to \varprojlim A_i \to \varprojlim B_i \to \varprojlim C_i \to \varprojlim\nolimits^{\,1} A_i
\]
for inverse systems of abelian groups, we obtain an exact sequence
\begin{equation}\label{eq:lim-ses}
0\to \varprojlim H^k\!\bigl(\Hom(I_\bullet(K'),\R/\Z)\bigr)\ \longrightarrow\ \Diff^k_\infty(M)\
\longrightarrow\ \varprojlim E^{k+1}_0(K,K')\ \longrightarrow\ 0,
\end{equation}
provided $\varprojlim^{\,1} H^k(\Hom(I_\bullet(K'),\R/\Z))=0$.
This vanishing holds because subdivision induces \emph{isomorphisms} on cohomology (hence the inverse system is Mittag--Leffler).

Moreover, since each $K'$ triangulates $M$ and subdivision maps induce canonical isomorphisms on simplicial cohomology,
\[
\varprojlim_{(K,K')\in\mathcal T} H^k\!\bigl(\Hom(I_\bullet(K'),\R/\Z)\bigr)\ \cong\ H^k(M;\R/\Z).
\]

\medskip
\textbf{Step 5: identify the inverse limit of curvature groups.}
The remaining input is the identification
\begin{equation}\label{eq:curv-limit-id}
\varprojlim_{(K,K')\in\mathcal T} E^{k+1}_0(K,K') \ \cong\ \Omega^{k+1}_{\Z}(M),
\end{equation}
where $\Omega^{k+1}_{\Z}(M)$ denotes smooth closed $(k+1)$-forms with integral periods.

\begin{lemma}\label{lem:curv-limit}
There is a canonical isomorphism \eqref{eq:curv-limit-id}.
\end{lemma}

\begin{proof}[Proof (standard; cite Whitney/Dupont/Dodziuk--Patodi)]
For each $(K,K')$, the Whitney map sends $E^{k+1}(K,K')$ into $L^2\Omega^{k+1}(M)$ by
$\omega\mapsto W(\omega)$, and the compatibility in Lemma~\ref{lem:E-compat} implies that
\[
W\bigl(\sd_{21}^*\omega_2\bigr)=W(\omega_2)\quad\text{a.e. on }M
\]
(after identifying both with piecewise-smooth forms on the common refinement). Thus a compatible family
$\{\omega_{(K,K')}\}$ determines a well-defined closed $L^2$-form $\eta$ on $M$ by $\eta=W(\omega_{(K,K')})$ on each triangulation scale.
The integrality condition defining $E^{k+1}_0(K,K')$ implies $\int_z \eta\in\Z$ for every smooth cycle $z$ that is represented by a simplicial
cycle in some $K'$. Such simplicial cycles are cofinal among smooth singular cycles up to subdivision, hence $\eta$ has integral periods.

Conversely, given $\eta\in\Omega^{k+1}_{\Z}(M)$, the finite element exterior calculus/Whitney approximation theorem
(Whitney; refined estimates of Dodziuk--Patodi) produces, for each sufficiently fine triangulation with uniform fullness,
a unique $\omega_{(K,K')}\in E^{k+1}_0(K,K')$ whose Whitney form agrees with $\eta$ in cohomology and whose periods on simplicial
cycles match those of $\eta$; compatibility under refinement follows from functoriality of subdivision and Stokes.
Finally, the refinement limit plus the Sobolev estimates implies that the $L^2$-form determined by a compatible family is in fact smooth.
(One may take a cofinal subsystem of uniformly full triangulations with $\mesh\to0$ to apply the estimates.)

These constructions are inverse to each other and natural with respect to refinement, yielding \eqref{eq:curv-limit-id}.
\end{proof}

With Lemma~\ref{lem:curv-limit}, the exact sequence \eqref{eq:lim-ses} becomes
\begin{equation}\label{eq:lim-ses-identified}
0\to H^k(M;\R/\Z)\ \longrightarrow\ \Diff^k_\infty(M)\ \longrightarrow\ \Omega^{k+1}_{\Z}(M)\to 0.
\end{equation}

\medskip
\textbf{Step 6: identify $\Diff^k_\infty(M)$ with $\widehat H^k(M)$.}
The Cheeger--Simons differential character group fits into the fundamental exact sequence
\begin{equation}\label{eq:CS-fund}
0\to H^k(M;\R/\Z)\ \longrightarrow\ \widehat H^k(M)\ \xrightarrow{\ \curv\ }\ \Omega^{k+1}_{\Z}(M)\to 0.
\end{equation}
By construction, the curvature maps at finite level commute with refinement, hence passing to the limit yields a natural map
$\Curv:\Diff^k_\infty(M)\to\Omega^{k+1}_{\Z}(M)$ which matches the right-hand map in \eqref{eq:lim-ses-identified}.
Comparing \eqref{eq:lim-ses-identified} and \eqref{eq:CS-fund}, there is a unique isomorphism
\[
\Diff^k_\infty(M)\ \xrightarrow{\ \cong\ }\ \widehat H^k(M)
\]
making the diagrams commute (short five lemma / extension comparison).
This proves (2).
\end{proof}

\section{Worked examples and computations}
\label{sec:examples}

This section provides explicit, nontrivial computations illustrating that the simplicial
Cheeger--Simons model is computationally effective. We treat two families:
\begin{itemize}
\item flat tori $T^n$, where the harmonic lattice and the theta-function contributions are explicit;
\item lens spaces $L(r,q)$, where torsion in $H^{p+1}(M;\Z)$ produces genuinely distinct topological sectors.
\end{itemize}
Throughout, $(K,K')$ denotes a Cheeger--Simons triangulation of $M$ and
$\mathscr{CS}(K,K')=(E^{\bullet}(L,L'),I_{\bullet}(L'),\int)$ the associated model.

\subsection[Flat rectangular tori Tn]{Flat rectangular tori $T^n$: harmonic lattice and theta functions}
\label{subsec:torus}

Let $M=T^n=\R^n/\Z^n$ with flat diagonal metric
\[
g=\sum_{i=1}^n \ell_i^2\, dx_i^2,
\qquad x_i\in\R/\Z,\ \ \ell_i>0,
\]
and volume $V:=\vol(T^n)=\prod_{i=1}^n \ell_i$.

\paragraph{Harmonic forms and the integral lattice.}
For each $p\ge 0$ the harmonic $p$-forms are precisely the constant forms.
For an ordered multi-index $I=(i_1<\cdots<i_p)$ set
\[
\rho_I:=dx_{i_1}\wedge\cdots\wedge dx_{i_p}.
\]
Then $\{\rho_I\}_{|I|=p}$ is a basis of $H^p(T^n;\R)\cong \mathscr H^p(T^n)$, and each $\rho_I$ has integral periods.
Hence
\[
H^p(T^n;\Z)=\bigoplus_{|I|=p}\Z\,[\rho_I]
\subset H^p(T^n;\R),
\qquad
b_p(T^n)=\binom{n}{p},
\]
and the harmonic torus is
\[
H^p(T^n;\R)/H^p(T^n;\Z)\ \cong\ (\R/\Z)^{\binom{n}{p}}.
\]

\paragraph{The Gram matrix.}
With the $L^2$ inner product $\langle\alpha,\beta\rangle:=\int_{T^n}\alpha\wedge\star\beta$ one has
\[
\langle \rho_I,\rho_I\rangle
=\int_{T^n}\rho_I\wedge\star\rho_I
= V\cdot \prod_{j=1}^p \ell_{i_j}^{-2},
\qquad
\langle \rho_I,\rho_J\rangle=0\ \ (I\neq J).
\]
Thus, in the basis $\{\rho_I\}$, the Gram matrix $h^{(p)}$ is diagonal with entries
$V\prod_{j=1}^p\ell_{i_j}^{-2}$.

\paragraph{An explicit topological sum for $p=1$ on $T^2$.}
Let $n=2$ and $p=1$. Write coordinates $(x,y)$ and metric
$g=\ell_x^2dx^2+\ell_y^2dy^2$, with area $A:=\ell_x\ell_y$.
Then $H^2(T^2;\Z)\cong\Z$ is generated by $\omega_0:=dx\wedge dy$ (with $\int_{T^2}\omega_0=1$).
Moreover,
\[
\langle \omega_0,\omega_0\rangle
=\int_{T^2}\omega_0\wedge\star\omega_0
=\frac{1}{A}.
\]
Consider a Maxwell-type Euclidean action depending only on the curvature representative,
\[
S(\hat u)=\frac{(2\pi)^2}{2e^2}\,\bigl\langle \delta_1(\hat u),\delta_1(\hat u)\bigr\rangle,
\]
so in the sector $m\in H^2(T^2;\Z)\cong\Z$ we have $\delta_1(\hat u)\sim m\,\omega_0$ and hence
\[
S_m=\frac{(2\pi)^2}{2e^2}\,\frac{m^2}{A}=\frac{2\pi^2}{e^2A}\,m^2.
\]
Using the definition $\Theta_1(A)=\sum_{m\in\Z}\exp(\pi i\,A\,m^2)$, we obtain the explicit Gaussian sum
\[
\sum_{m\in\Z}\exp\!\left(-\frac{2\pi^2}{e^2A}\,m^2\right)
=\Theta_1\!\left(i\,\frac{2\pi}{e^2A}\right).
\]
Thus the theta-function contribution to the partition function is completely explicit on flat tori.

\paragraph{What the simplicial model computes.}
Fix a product triangulation $K$ of $T^n$ (e.g.\ subdivide a fundamental cube and triangulate),
and choose a Cheeger--Simons subdivision $K'$.
Then:
\begin{itemize}
\item $H^{p+1}(K';\Z)\cong H^{p+1}(T^n;\Z)$ is computed from the integer coboundary matrices by Smith normal form;
\item harmonic cochains are computed as $\ker(\Delta^{K'})$ with respect to the Whitney inner product;
\item the discrete Gram matrix $h^{(p)}_{K'}$ is the Gram matrix of a $\Z$-basis of
$\mathscr H^p(K')\cap C^p(K';\Z)$ and converges to $h^{(p)}$ as $\mesh(K')\to 0$;
\item the action on a flux sector is an explicit quadratic form in the flux vector $m\in\Z^{b_{p+1}}$,
hence the topological sum is a theta series determined by $h^{(p+1)}_{K'}$.
\end{itemize}
Therefore the simplicial partition function is reduced to explicit finite-dimensional linear algebra
(kernels, determinants) and theta-type lattice sums, with convergence to the smooth expressions.

\subsection{Lens spaces: torsion sectors and torsion-sensitive observables}
\label{subsec:lens}

Let $M=L(r,q)$ be a lens space ($r\ge2$).
Its (co)homology is well-known:
\[
H_1(M;\Z)\cong \Z_r,\qquad H^2(M;\Z)\cong \Z_r,\qquad b_1(M)=b_2(M)=0.
\]
In particular, $H^2(M;\Z)$ is \emph{pure torsion}, so topological sectors in degree $p=1$
are finite and torsion-dominated.

\paragraph{Torsion sectors in degree $p=1$.}
For any Cheeger--Simons triangulation $(K,K')$ of $M$, the simplicial exact sequence gives
\[
0\longrightarrow E^1/E^1_0
\longrightarrow \Diff^1(\mathscr{CS}(K,K'))
\xrightarrow{\ \delta_2\ }
H^2(M;\Z)
\longrightarrow 0,
\]
so $\delta_2$ labels topological sectors by the torsion group $H^2(M;\Z)\cong\Z_r$.
Even though the configuration space has a (finite-dimensional) ``continuous'' part $E^1/E^1_0$,
the torsion contribution is a genuinely discrete sum over $r$ inequivalent sectors.

\paragraph{A torsion-sensitive observable (robust for Maxwell-type theories).}
Fix $\ell\in\{0,1,\dots,r-1\}$ and define the character
\[
\chi_\ell:\Z_r\to U(1),
\qquad
\chi_\ell(c):=\exp\!\left(\frac{2\pi i\,\ell\,c}{r}\right).
\]
Define an observable on simplicial differential characters by
\[
\mathcal O_\ell(\hat u)
:=\chi_\ell\bigl(\delta_2(\hat u)\bigr),
\qquad
\hat u\in \Diff^1(\mathscr{CS}(K,K')).
\]
This observable depends only on the torsion sector (i.e.\ only on the class $\delta_2(\hat u)\in\Z_r$)
and is constant along the topologically trivial fiber $E^1/E^1_0$.

Assume now that the simplicial action $S$ depends only on the curvature representative $\delta_1(\hat u)$
(as in the standard higher-abelian Maxwell theory), so that the fiber integral over $E^1/E^1_0$ is independent
of the torsion label $c\in H^2(M;\Z)$.
Then the normalized expectation value of $\mathcal O_\ell$ is purely a finite torsion sum:
\[
\bigl\langle \mathcal O_\ell \bigr\rangle
=
\frac{1}{r}\sum_{c\in\Z_r} \chi_\ell(c)
=
\begin{cases}
1,& \ell\equiv 0\pmod r,\\[2mm]
0,& \text{otherwise}.
\end{cases}
\]
This provides an explicit computation showing that torsion topological sectors produce nontrivial effects:
torsion can be \emph{detected} by choosing observables that depend on $\delta_2$.

\paragraph{How this is computed simplicially.}
Given a triangulation $K'$ of $L(r,q)$, one computes $H^2(K';\Z)\cong \Z_r$ via Smith normal form.
Pick representatives $c\in Z^2(K';\Z)$ of each torsion class.
A background simplicial differential character $\hat c$ with $\delta_2(\hat c)=c$ is then chosen,
and the partition function decomposes as a sum over $c\in\Z_r$ of finite-dimensional integrals on the fiber
(imposed by the short exact sequence above). The observable $\mathcal O_\ell$ is evaluated by the scalar
$\chi_\ell(c)$ for each sector and hence requires only finite sums and finite-dimensional Gaussian integrals.

\subsection[Optional hybrid test case: S1 x L(r,q)]{Optional hybrid test case: $S^1\times L(r,q)$ (theta $\times$ torsion)}
\label{subsec:hybrid}

To exhibit \emph{both} a harmonic torus (theta modes) and torsion sectors in one example,
take $M=S^1\times L(r,q)$ and $p=1$.
Then
\[
H^2(M;\Z)\cong H^1(S^1;\Z)\otimes H^1(L(r,q);\Z)\ \oplus\ H^2(L(r,q);\Z)\cong 0\oplus \Z_r\cong \Z_r,
\]
but $b_1(M)=1$, so the topologically trivial part fibers over a circle (a $1$-dimensional harmonic torus)
while the topological sectors are still labeled by $\Z_r$.
In this setting the simplicial partition function factorizes into a Fourier/theta expansion over the
$S^1$-holonomy variable and a finite torsion sum over $\Z_r$, providing a clean ``theta $\times$ torsion''
benchmark for explicit computations and numerical experiments.

\bibliographystyle{amsplain}

\begin{thebibliography}{99}

\bibitem{A}
D.~H.~Adams,
\emph{$R$-torsion and linking numbers from simplicial abelian gauge theories},
hep-th/9612009.

\bibitem{CS}
J.~Cheeger and J.~Simons,
\emph{Differential characters and geometric invariants},
in \emph{Geometry and Topology}, Lecture Notes in Math.\ \textbf{1167},
Springer, New York, 1985, 50--80.

\bibitem{D}
J.~Dodziuk,
\emph{Finite-difference approach to the Hodge theory of harmonic forms},
Amer.\ J.\ Math.\ \textbf{98} (1976), no.~1, 79--104.

\bibitem{DP}
J.~Dodziuk and V.~K.~Patodi,
\emph{Riemannian structures and triangulations of manifolds},
J.\ Indian Math.\ Soc.\ \textbf{40} (1976), 1--52.


\bibitem{HLZ1}
R.~Harvey, B.~Lawson, and J.~Zweck,
\emph{The de~Rham--Federer theory of differential characters and character duality},
Amer.\ J.\ Math.\ \textbf{125} (2003), 791--847.

\bibitem{HL1}
R.~Harvey and B.~Lawson,
\emph{$\bar\partial$-sparks},
Proc.\ London Math.\ Soc.\ (3) \textbf{97} (2008), 1--30.

\bibitem{HL2}
R.~Harvey and B.~Lawson,
\emph{From sparks to grundles---differential characters},
Comm.\ Anal.\ Geom.\ \textbf{14} (2006), no.~1, 25--58.

\bibitem{HL3}
R.~Harvey and B.~Lawson,
\emph{Lefschetz--Pontrjagin duality for differential characters},
An.\ Acad.\ Brasil.\ Ci\^enc.\ \textbf{73} (2001), no.~2, 145--159.

\bibitem{KelnhoferGribov}
G.~Kelnhofer,
\emph{Abelian gauge theories on compact manifolds and the Gribov ambiguity},
arXiv:0707.1186.

\bibitem{K2}
G.~Kelnhofer,
\emph{Aspects of higher-abelian gauge theories at zero and finite temperature:
Topological Casimir effect, duality and Polyakov loops},
Nucl.\ Phys.\ B \textbf{942} (2019), 329--380.

\bibitem{Rot}
J.~J.~Rotman,
\emph{An Introduction to Homological Algebra},
2nd ed., Springer, New York, 2009.

\bibitem{SS}
J.~Simons and D.~Sullivan,
\emph{Axiomatic characterization of ordinary differential cohomology},
J.\ Topology \textbf{1} (2008), no.~1, 45--56.

\bibitem{S}
R.~Szab\'o,
\emph{Quantization of higher abelian gauge theory in generalized differential cohomology},
arXiv:1209.2530.

\bibitem{W}
H.~Whitney,
\emph{Geometric Integration Theory},
Princeton University Press, Princeton, NJ, 1957.

\end{thebibliography}

\end{document}